\documentclass[letter]{aa} 

\usepackage{graphicx}
\usepackage{txfonts}
\usepackage{hyperref}
\hypersetup{
     colorlinks = true,
     linkcolor = blue,
     anchorcolor = blue,
     citecolor = blue,
     filecolor = blue,
     urlcolor = blue
     }
\usepackage{pdflscape} 
\usepackage{afterpage} 
\defcitealias{postijstar}{P18}
\defcitealias{paperIBFR}{MP21}
\defcitealias{fall2018}{FR18}
\defcitealias{obreschkow2016}{O16}

\usepackage{tabularx}
\usepackage{amsmath}
\begin{document} 

   \title{A tight angular-momentum plane for disc galaxies}
   \titlerunning{A tight angular-momentum plane for disc galaxies}
\authorrunning{Mancera Pi\~na et al.}

   \author{Pavel E. Mancera Pi\~na\inst{1,2}\fnmsep\thanks{\email{pavel@astro.rug.nl}},
          Lorenzo Posti\inst{3}, Gabriele Pezzulli\inst{1}, Filippo Fraternali\inst{1}, S. Michael Fall\inst{4}, Tom Oosterloo\inst{2,1} and Elizabeth A. K. Adams\inst{2,1}
          }
   \institute{Kapteyn Astronomical Institute, University of Groningen, Landleven 12, 9747 AD Groningen, The Netherlands
         \and
       ASTRON, Netherlands Institute for Radio Astronomy, Oude Hoogeveensedijk 4, 7991 PD, Dwingeloo, The Netherlands
        \and 
        Observatoire astronomique de Strasbourg, Universit\'e de Strasbourg, 11 rue de l'Universit\'e, 67000 Strasbourg, France  
        \and Space Telescope Science Institute, 3700 San Martin Drive, Baltimore, MD 21218, USA        }


 
  \abstract{The relations between the specific angular momenta ($j$) and masses ($M$) of galaxies are often used as a benchmark in analytic models and hydrodynamical simulations as they are considered to be amongst the most fundamental scaling relations. Using accurate measurements of the stellar ($j_\ast$), gas ($j_{\rm gas}$), and baryonic ($j_{\rm bar}$) specific angular momenta for a large sample of disc galaxies, we report the discovery of tight correlations between $j$, $M$, and the cold gas fraction of the interstellar medium ($f_{\rm gas}$). At fixed $f_{\rm gas}$, galaxies follow parallel power laws in 2D $(j,M)$ spaces, with gas-rich galaxies having a larger $j_\ast$ and $j_{\rm bar}$ (but a lower $j_{\rm gas}$) than gas-poor ones. The slopes of the relations have a value around 0.7. These new relations are amongst the tightest known scaling laws for galaxies. In particular, the baryonic relation ($j_{\rm bar}-M_{\rm bar}-f_{\rm gas}$), arguably the most fundamental of the three, is followed not only by typical discs but also by galaxies with extreme properties, such as size and gas content, and by galaxies previously claimed to be outliers of the standard 2D $j-M$ relations. The stellar relation ($j_{\ast}-M_{\ast}-f_{\rm gas}$) may be connected to the known $j_\ast-M_\ast-$bulge fraction relation; however, we argue that the $j_{\rm bar}-M_{\rm bar}-f_{\rm gas}$ relation can originate from the radial variation in the star formation efficiency in galaxies, although it is not explained by current disc instability models.}


   \maketitle
%

\section{Introduction}
\label{sec:intro}
Despite the remarkable diversity of galaxy properties observed in the present-day Universe, a number of physical parameters of galaxies appear to correlate with one another and form tight scaling laws. Such relations are of paramount importance in our quest to understand galaxy formation and evolution (e.g. \citealt{tullyfisher,fall83, burstein1997,mcgaugh2000,marconi_blackholes,cappellari_atlasXV,wang16}). 

\noindent
Since early models of galaxy formation were proposed, it has become clear that mass and angular momentum are two fundamental parameters controlling the evolution of galaxies (e.g. \citealt{fall1980,dalcanton_discs,mo98}). From an observational point of view, starting from the work by \citet{fall83}, different authors have characterised the scaling relation between stellar mass ($M_\ast$) and stellar specific angular momentum ($j_\ast = J_\ast/M_\ast$, where $J_\ast$ is the angular momentum) as the $j_\ast-M_\ast$ or Fall relation (e.g. \citealt{romanowsky, OG14, postijstar}). This $j_\ast-M_\ast$ law has been widely used in recent years to constrain and test both (semi-)analytic models and hydrodynamical simulations (e.g. \citealt{genel2015,pedrosa2015,obreja2016,lagos_eagle,tremmel2017,elbadry18,stevens2018,zoldan18,irodotou2019}).

\citet[][hereafter \citetalias{paperIBFR}]{paperIBFR}, derived accurate measurements of the stellar ($j_\ast$), (cold neutral) gas ($j_{\rm gas}$), and baryonic ($j_{\rm bar}$) specific angular momentum for a large sample of irregular spiral and dwarf  galaxies (see also e.g. \citealt{OG14,kurapati}). They determined the $j-M$ relations for the three components and fitted them with unbroken power laws. They also noticed that the residuals from the best fitting relations correlate with the gas fraction ($f_{\rm gas}=M_{\rm gas}/M_{\rm bar}$, with $M_{\rm gas}$ and $M_{\rm bar}$ the gas and baryonic masses, respectively). These trends, also seen in a few semi-analytic models (e.g. \citealt{stevens2018,zoldan18}), may indicate that the gas content plays an important role in the $j-M$ relations. In this letter we build upon that result and report the discovery of new and very tight correlations between mass, specific angular momentum, and gas fraction. We show that disc galaxies across $\sim4$ orders of magnitude in mass lie in very tight planes in the ($j,M,f_{\rm gas}$) spaces.

\section{Definition of $j$ and galaxy sample}
\label{sec:background}
The stellar and gas specific angular momenta of a galaxy are defined as   
\begin{equation}
    j_{i}(<R) = \dfrac{\int_0^{R} R'^{2} ~\Sigma_{i}(R')~V_{i}(R')~dR'}{\int_0^{R}  R'~\Sigma_{i}(R')~dR'}~,
    \label{eq:j}
\end{equation}
\noindent
with $R$ being the galactocentric cylindrical radius, $\Sigma_{i}$ the stellar or gas face-on surface density, and $V_{i}$ the stellar or gas rotation velocity. Then, $j_\ast$ and $j_{\rm gas}$ can be combined to obtain
\begin{equation}
\label{eq:jbar}
    j_{\rm bar} = f_{\rm gas} j_{\rm gas} + (1-f_{\rm gas})j_\ast~.
\end{equation}
For $f_{\rm gas} = M_{\rm gas}/M_{\rm bar}$, we assumed $M_{\rm bar} = M_\ast + M_{\rm gas}$, with $M_{\rm gas} = 1.33M_{\rm HI}$, where $M_{\rm HI}$ is the mass of neutral atomic hydrogen and the factor 1.33 accounts for the presence of helium. While we neglected any contribution from molecular gas, in Appendix~\ref{sec:appH2} we show that its inclusion does not change the results found in this letter.\footnote{Our `gas' refers only to the interstellar medium, and there is no attempt to include the largely unconstrained contribution of the gas outside galaxy discs. Although the sum of our gas and stars does not represent the `whole' baryonic budget of a galaxy, we prefer to keep this nomenclature for the sake of consistency with the literature.}

\citetalias{paperIBFR} compiled a high-quality sample of 157 nearby galaxies, predominantly discs. All the galaxies have near-IR photometry and extended HI rotation curves, allowing their stellar discs and rotation velocities to be traced robustly. The sample includes dwarf and massive galaxies, spanning the mass range $7 \lesssim$~log($M_\ast$/$M_\odot) \lesssim 11.5$, $6 \lesssim$~log($M_{\rm gas}$/$M_\odot) \lesssim 10.5$, $8 \lesssim$~log($M_{\rm bar}$/$M_\odot) \lesssim 11.5$, and with $0.01<f_{\rm gas}<0.97$ and a typical relative uncertainty, $\delta {f_{\rm gas}}/f_{\rm gas} \approx 0.2$~dex (median $\delta {f_{\rm gas}} \approx 0.05$). While not complete, the sample is representative of the population of regularly rotating nearby discs, like other large samples commonly used in the literature (e.g. \citealt{sparc,anastasia1}). Using the near-IR photometry and HI rotation curves, \citetalias{paperIBFR} built cumulative radial profiles for $j_\ast$ (applying a correction to convert $V_{\rm gas}$ into $V_\ast$), $j_{\rm gas}$, and $j_{\rm bar}$. By selecting only galaxies with radially convergent measurements of angular momentum, they built a sample of 130, 87, and 106 galaxies with accurate $j_\ast$, $j_{\rm gas}$, and $j_{\rm bar}$, respectively. For more details we refer the reader to \citetalias{paperIBFR}, and in \href{https://unishare.nl/index.php/s/NMQRYfrrDpj8iaj}{this link} we provide online tables listing $j$, $M$, $f_{\rm gas}$, distance, and Hubble type for the galaxy sample.

\citetalias{paperIBFR} fitted the $j-M$ relations with power laws of the form
\begin{equation}
\label{eq:linearmodel}
     \log \left(\dfrac{j_i}{\rm{kpc\ km\ s^{-1}}} \right) = m_i[\log(M_{i}/M_\odot)-10] + n_i~,
\end{equation}
with the subscript $i$ representing the stellar, gaseous, or baryonic component. The best fitting power laws have slopes $m_i$ of about 0.5, 1.0, and 0.6 for stars, gas, and baryons, respectively, and an intrinsic scatter of 0.15~dex.

\section{The $j-M-f_{\rm gas}$ planes}
\label{sec:thespaces}
\subsection{Best fitting planes}

\citetalias{paperIBFR} (see their Fig.~7) also found systematic trends with $f_{\rm gas}$ in the residuals of the three $j-M$ laws. To see if introducing a dependence of the $j-M$ laws on $f_{\rm gas}$ can explain these trends, we fitted the $(j,M,f_{\rm gas})$ data with planes. We fitted the data points with the model
\begin{equation}
    \log \left(\dfrac{j_{i}}{\rm{kpc\ km\ s^{-1}}} \right) = \alpha_{i}\log(M_{i}/M_\odot) + \beta_{i} \log(f_{\rm gas})~ + \gamma_{i}~.
    \label{eq:3dplanes}
\end{equation}
Therefore, we assumed that, in contrast to Eq.~\ref{eq:linearmodel}, $j_i$ also depends on $f_{\rm gas}$.\footnote{Since our planes depend on $\log(f_{\rm gas})$, they become hard to interpret when $f_{\rm gas} \rightarrow 0$, preventing us from making extrapolations for galaxies with $f_{\rm gas} < 0.01$. Using $f_{\rm gas}$ instead of $\log(f_{\rm gas})$ produces less satisfactory fits when compared to the observations, so we prefer $\log(f_{\rm gas})$ despite its limitations when $f_{\rm gas} \rightarrow 0$. We also note that, given Eq.~\ref{eq:jbar}, the $j-M$ relations cannot all be exactly planes. However, fitting planes in the ($j,M,f_{\rm gas}$) spaces is a very useful empirical approach.} We performed the fit using the \textsc{r} package \textsc{hyper-fit} \citep{hyperfit}, including a term for the intrinsic scatter $\sigma_\perp$. We assumed log-normal uncertainties in $j, M$, and $f_{\rm gas}$, and, using a Monte Carlo sampling method, we took into account the fact that uncertainties in the distance, inclination, and mass-to-light ratio of a given galaxy drive correlated uncertainties (also provided in our electronic tables) between $\log(M), \log(j)$, and $\log(f_{\rm gas})$. We stress that taking these correlations into account is important: Neglecting them can artificially lower the intrinsic scatter of the planes by a factor of two to three. 

The best fitting coefficients are reported in Table~\ref{tab:coeff_planes}. The orthogonal intrinsic scatter of our best fitting planes is significantly smaller than for the 2D relations. The log-marginal likelihood is also higher (i.e. better) for the 3D planes: by 27, 40, and 43 units for stars, gas, and baryons, respectively. We conclude that the inclusion of $f_{\rm gas}$ into the $j-M$ laws is statistically meaningful. 

\begin{table}
\caption{Coefficients of the best fitting $j-M-f_{\rm gas}$ planes.}
\label{tab:coeff_planes}
\begin{center} 
\begin{tabular}{ccccc}
        \hline \noalign{\smallskip}
 & $\alpha$ & $\beta$ & $\gamma$ & $\sigma_\perp$ \\ \noalign{\smallskip}
   \hline \noalign{\smallskip}
   Stars & 0.67 $\pm$ 0.03 & 0.51 $\pm$ 0.08 & -3.62 $\pm$ 0.23 & 0.10 $\pm$ 0.01 \\ \noalign{\smallskip}
    Gas & 0.78 $\pm$ 0.03 & -0.49 $\pm$ 0.04 & -4.64 $\pm$ 0.25 & 0.08 $\pm$ 0.01 \\ \noalign{\smallskip}
    Baryons & 0.73 $\pm$ 0.02 & 0.46 $\pm$ 0.05 & -4.25 $\pm$ 0.19 & $0.08 \pm 0.01$ \\ \noalign{\smallskip}
    \hline
\end{tabular}
  \end{center}
\end{table}

\begin{figure*}
    \centering
    \includegraphics[scale=0.41]{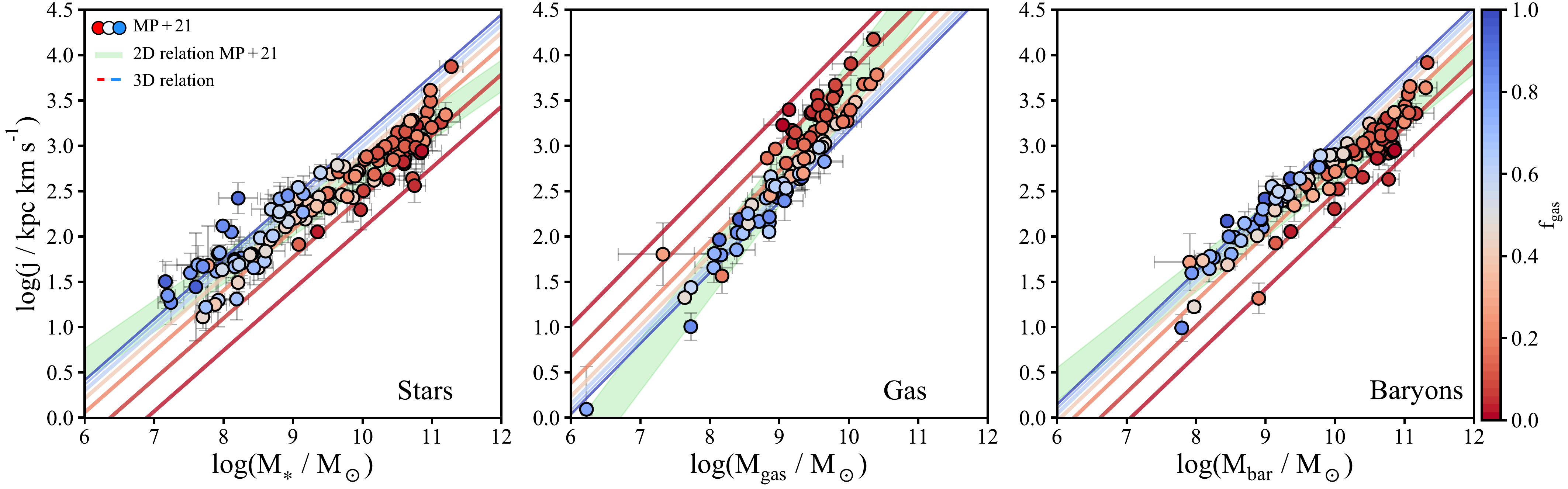}
    \caption{Stellar, gas, and baryonic $j-M-f_{\rm gas}$ planes, projected into the 2D $(j,M)$ spaces. Galaxies are colour-coded according to their $f_{\rm gas}$ and are compared with lines of constant $f_{\rm gas}$ according to Eq.~\ref{eq:3dplanes} and the best fitting coefficients of Table~\ref{tab:coeff_planes}. From red to blue, the lines are at $f_{\rm gas} = 0.01, 0.05, 0.2, 0.4, 0.6, 0.8, 1$. For comparison, we show in green the best fitting 2D $j-M$ relations from \citetalias{paperIBFR} and their intrinsic scatter.}
    \label{fig:3panels}
\end{figure*}

In Fig.~\ref{fig:3panels} we compare the observed distribution of galaxies with our three best fitting planes; the figure shows the 3D $j-M-f_{\rm gas}$ planes projected into the 2D $(j,M)$ spaces. Galaxies are colour-coded according to their $f_{\rm gas}$, and we overlay our lines of constant $f_{\rm gas}$ derived from our best fitting planes. The fits provide a very good description of the data, in line with the low intrinsic scatter we find for all the planes. 
By construction, the three $j-M-f_{\rm gas}$ planes are characterised by their $M$ slopes ($\alpha_{\ast}$, $\alpha_{\rm gas}$, and $\alpha_{\rm bar}$) and $f_{\rm gas}$ slopes ($\beta_{\ast}$, $\beta_{\rm gas}$, and $\beta_{\rm bar}$). Projected into the ($j,M$) spaces, the $f_{\rm gas}$ slopes act as a normalisation for $j$. At fixed $M_\ast$ ($M_{\rm bar}$), gas-rich galaxies have a higher $j_\ast$ ($j_{\rm bar}$) than gas-poor ones, while gas-poor galaxies show a higher $j_{\rm gas}$. For stars and baryons, the 3D relations become steeper ($\alpha_{\ast}=0.67\pm 0.03, \alpha_{\rm{bar}}=0.73\pm 0.03$) than the 2D ones from \citetalias{paperIBFR} ($m_{\ast}=0.53\pm 0.02, m_{\rm{bar}}=0.60\pm 0.02$) once $f_{\rm gas}$ is taken into account, while the slope of the gas relation becomes shallower ($\alpha_{\rm{gas}}=0.78\pm 0.03, m_{\rm{gas}}=1.02\pm 0.04$). 
Given the different coefficients, the 2D $j-M$ relations (shown in Fig.~\ref{fig:3panels} as green bands) differ from the projection of the 3D planes in the $(j,M)$ spaces, especially at $M<10^8 M_\odot$.

A remarkable property of our new scaling laws is their low intrinsic scatter. Given that the baryonic $j_{\rm bar}-M_{\rm bar}-f_{\rm gas}$ plane incorporates the stellar and cold gas components, we argue that this is likely the most fundamental of the three relations, although its intrinsic scatter is similar to the relation for the gas. Very few other scaling laws are thought to have a comparably low intrinsic scatter, for instance the HI mass-size relation \citep{wang16} or the baryonic Tully-Fisher relation (BTFR; \citealt{mcgaugh2000,anastasiaphotometry}). In fact, our baryonic plane can in principle be used as a distance estimator, with an uncertainty $\delta D/D = (\delta j_{\rm bar}/j_{\rm bar})/\vert 2\alpha_{\rm bar} -1\vert$ at fixed $M_{\rm bar}$.

\subsection{The similarities of the $\alpha$ slopes}
\label{sec:similarslopes}
\begin{figure}
    \centering
    \includegraphics[scale=0.42]{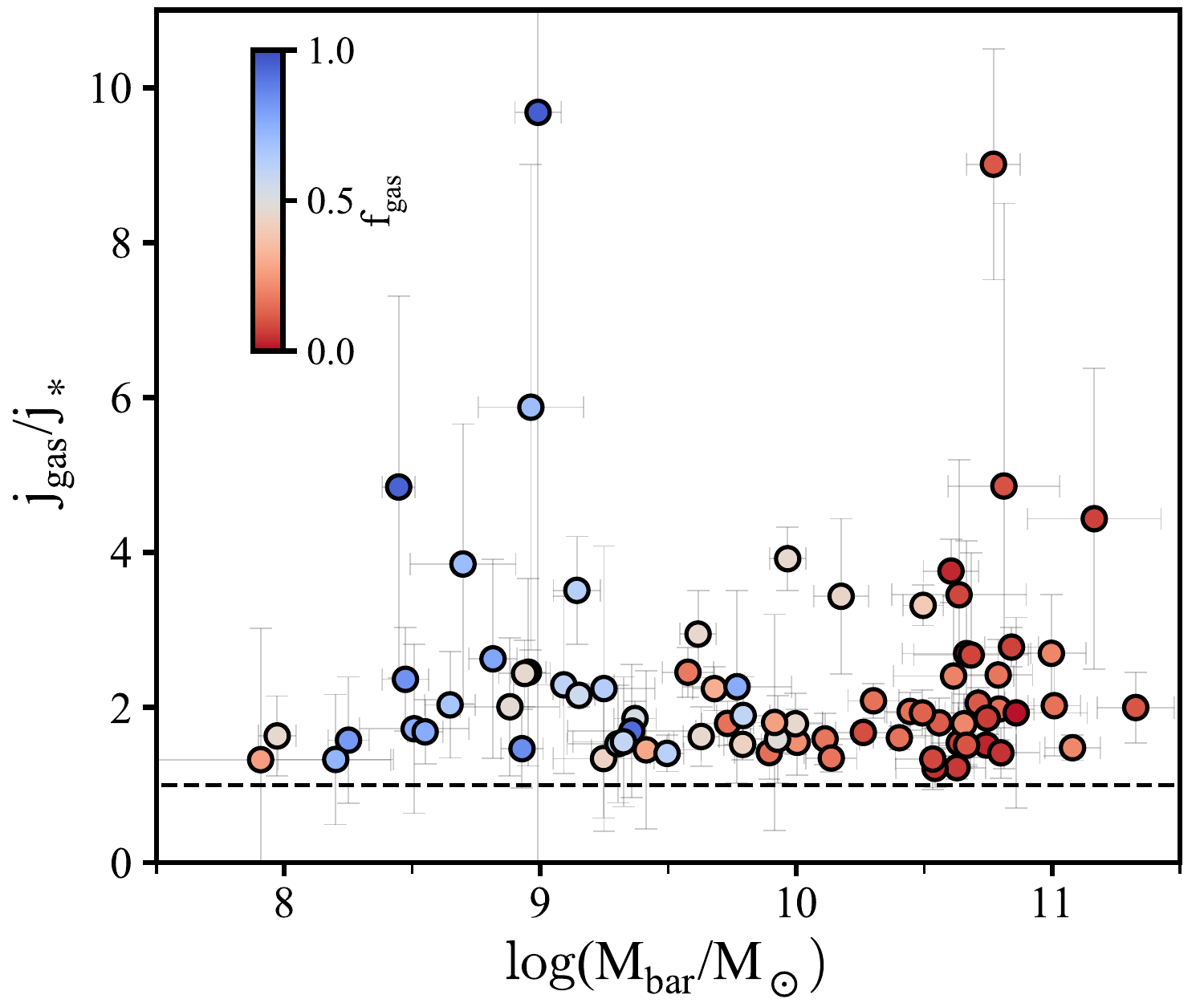}
    \caption{ $j_{\rm gas}/j_\ast$ ratio as a function of $M_{\rm bar}$. Galaxies are colour-coded by their $f_{\rm gas}$, and the dashed black line corresponds to $j_{\rm gas}/j_\ast=1$. Our galaxies (those with convergent $j_\ast$ and $j_{\rm gas}$ from \citetalias{paperIBFR}) cluster at $j_{\rm gas}/j_\ast \sim 2$ at all $M_{\rm bar}$, albeit with a significant scatter.}
    \label{fig:ratio}
\end{figure}

The three $\alpha$ slopes of our $j-M-f_{\rm gas}$ planes are relatively close to one another and to the value $2/3$ expected for their parent dark matter halos \citep{fall83}, which suggests some degree of structural self-similarity between different baryonic components and the dark matter halo. From a mathematical point of view, if we rewrite Eq.~\ref{eq:jbar} in terms of $j_\ast = B M_{\ast}^{\alpha_{\ast}}$ (with $B$ a function that depends only on $f_{\rm gas}$) and $M_\ast = M_{\rm bar} (1-f_{\rm gas})$, we obtain
\begin{equation}
    j_{\rm bar} = B~(1-f_{\rm gas})^{\alpha_{\ast}} \left[\dfrac{j_{\rm gas}}{j_\ast}f_{\rm gas} + (1-f_{\rm gas})  \right] M_{\rm bar}^{\alpha_{\ast}}~.
\end{equation}
In a similar way, considering now $j_{\rm gas} = C M_{\rm gas}^{\alpha_{\rm gas}}$ (with $C$ a function that depends only on $f_{\rm gas}$), we find
\begin{equation}
    j_{\rm bar} = C~f_{\rm gas}^{\alpha_{\rm gas}} \left[\dfrac{j_{\ast}}{j_{\rm gas}}(1-f_{\rm gas}) + f_{\rm gas}  \right] M_{\rm bar}^{\alpha_{\rm gas}}~.
\end{equation}
Therefore, at fixed $f_{\rm gas}$, the slope $\alpha_{\rm bar}$ of the baryonic $j-M-f_{\rm gas}$ plane is expected to be similar to $\alpha_{\ast}$ and $\alpha_{\rm gas}$, provided that the ratio $j_{\rm gas}/j_\ast$ is independent of $M_{\rm bar}$. As shown in Fig.~\ref{fig:ratio}, for our sample, $j_{\rm gas}/j_\ast$, which is always larger than 1 and mostly within a narrow range (the $16^{\rm th}$ and $84^{\rm th}$ percentiles are 1.5 and 3.2, respectively), does not seem to correlate with $M_{\rm bar}$, in line with the near-parallelism of the three relations shown in Fig.~\ref{fig:3panels}. 

It is worth noticing that the $j_{\rm gas}/j_\ast$ ratio can be related to the relative extent of some characteristic size of the gaseous ($R_{\rm gas}$) and stellar ($R_\ast$) components of galactic discs, given that $j_{\rm gas}/j_\ast \approx R_{\rm gas}V_{\rm gas}/(R_\ast V_\ast) \approx R_{\rm gas}/R_\ast$. Although this is just an approximation, it can be useful in the physical interpretation of the $j-M-f_{\rm gas}$ relations (e.g. Sect.~\ref{sec:SFE}). In Appendix~\ref{sec:app_ratio} we show the expected dependence of $j_{\rm gas}/j_\ast$ on $M_{\rm bar}$ and $f_{\rm gas}$ derived from our best fitting $j_\ast-M_\ast-f_{\rm gas}$ and $j_{\rm gas}-M_{\rm gas}-f_{\rm gas}$ planes.

\subsection{No outliers of the baryonic $j-M-f_{\rm gas}$ law}
\begin{figure*}
    \centering
    \includegraphics[scale=0.51]{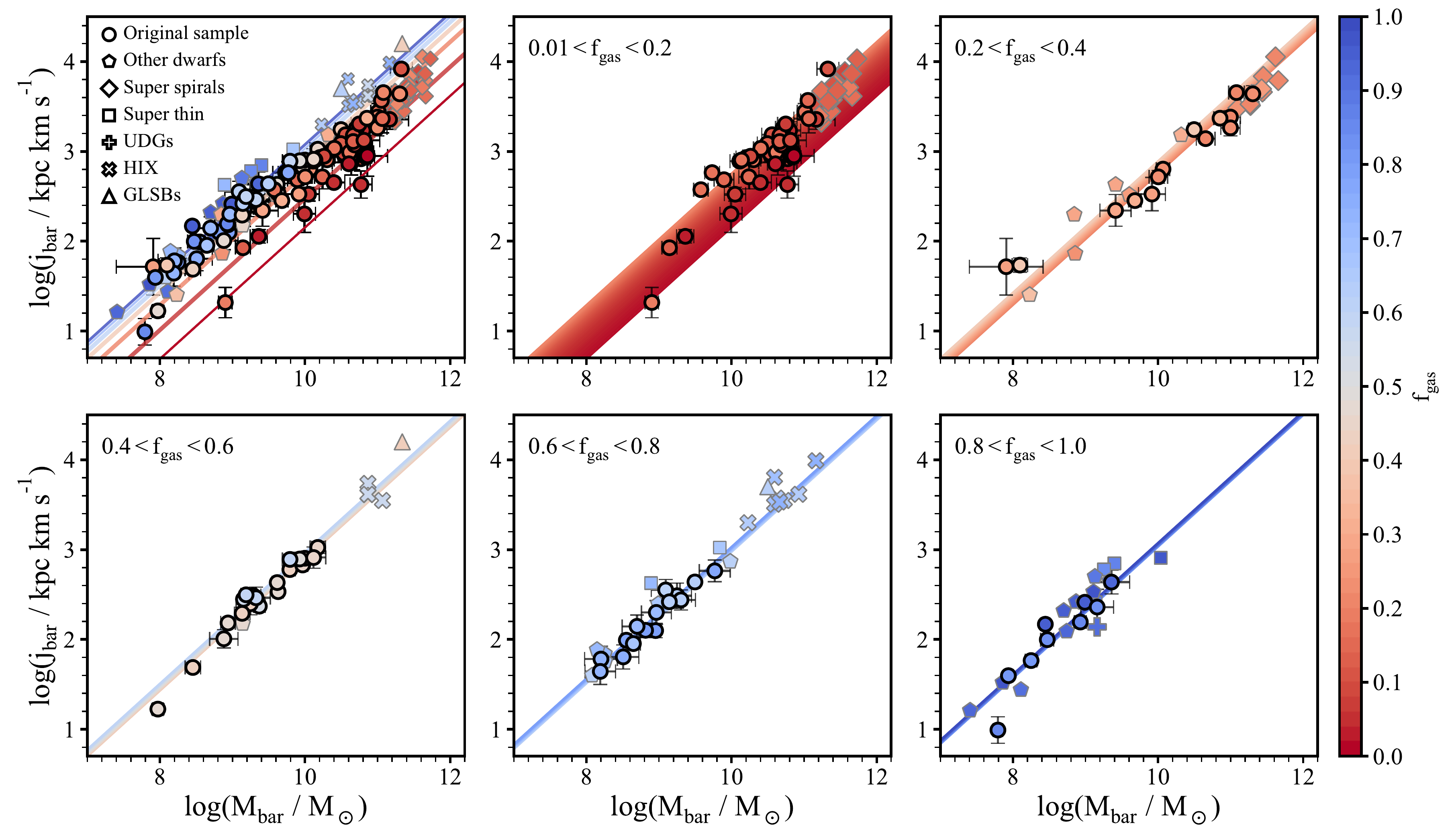}
    \caption{Baryonic $j_{\rm bar}-M_{\rm bar}-f_{\rm gas}$ plane for our original sample (circles) and a set of extreme galaxies (see text). The top left panel shows the relation for all the galaxies, while the remaining panels show the galaxies in bins of $f_{\rm gas}$ (given in the top left corner of each panel). On the first panel, the lines of constant $f_{\rm gas}$ are as in Fig.~\ref{fig:3panels}. In the remaining panels, the coloured areas enclose the region delimited by the whole $f_{\rm gas}$ bin. We remark that the coloured lines of our plane are derived by fitting only our data. The rest of the galaxies closely follow our fit.}
    \label{fig:jbarmbar_fgas}
\end{figure*}

In Fig.~\ref{fig:jbarmbar_fgas} we plot again our baryonic plane, this time splitting the galaxies into bins of $f_{\rm gas}$. In each panel we plot lines covering the whole range of $f_{\rm gas}$ within that bin. This allows the tightness of our baryonic relation to be appreciated in more detail.

We now investigate whether any objects from known galaxy populations could be outliers of our baryonic plane. We do this by comparing our relation -- derived using only our sample -- with galaxies from the literature that have been argued to be outliers of the 2D $j-M$ relations or that could be outliers given their extreme properties in size, $f_{\rm gas}$, or rotation velocity.

First, we considered a set of dwarf galaxies from the literature, specifically those from \citet{elson} and \citet{kurapati}, that do not overlap with our sample. It was claimed by those authors that some of these galaxies are off the 2D $j_{\rm bar}-M_{\rm bar}$ relation. We also included the dwarf `super thin' galaxies of \citet{superthin}, which have very high axis ratios and have been suggested to have higher $j_\ast$ than other dwarfs. Next, we looked at the `HI extreme galaxies' (HIXs) of \citet{lutz2018}, which have a particularly high $f_{\rm gas}$ for their $M_\ast$ and are claimed to have higher $j_{\rm bar}$ than average. Moreover, we added a sample of `super spiral' galaxies \citep{superspirals_enrico}, which are very large discs with masses a factor of ten larger than $L_\ast$ galaxies and were also claimed to be outliers of the $j_\ast-M_\ast$ relation \citep{superspirals}.\footnote{Since $j_{\rm gas}$ is not available for the super spirals, we assumed $j_{\rm gas}=1.9 j_\ast$, the median ratio found in the rest of our sample. This has little to no impact given their low gas content ($f_{\rm gas} \approx 0.15$).} Finally, we included the ultra-diffuse galaxies (UDGs) AGC~114905 and AGC~242019 (Mancera Pi\~na et al. in prep; \citealt{shi2021}) and the giant low surface brightness galaxies (GLSBs) Malin~1 and NGC~7589 \citep{lelli_glsb}. The UDGs are found to be outliers of the BTFR \citep{huds2019,huds2020}, and both UDGs and GLSBs are extreme galaxies with very extended discs for their $M_\ast$. With the caveat that some of these galaxies have larger uncertainties than our sample, given the different data quality, we added all these sets of galaxies into the $j_{\rm bar}-M_{\rm bar}-f_{\rm gas}$ plane in Fig.~\ref{fig:jbarmbar_fgas}. Remarkably, the location of all of these galaxies, given their $f_{\rm gas}$, is in very good agreement with the expectation of our scaling relation. We conclude that even extreme galaxies such as HIXs, UDGs, and GLSBs obey the $j_{\rm bar}-M_{\rm bar}-f_{\rm gas}$ law.

\section{Discussion}
\label{sec:interpretation}

\subsection{Stellar relation: The link with bulge fraction}
Previous works (e.g. \citealt{fall83,romanowsky,cortese2016,fall2018}) found that the relation between $j_\ast$ and $M_\ast$ depends on the bulge-to-total mass fraction, $\mathcal{B}_\ast$. 
\citet[][hereafter \citetalias{fall2018}]{fall2018}, proposed a model in which discs and spheroids follow relations of the form $j_\ast \propto M_\ast^{0.67}$ but with different normalisation, with spheroids shifting downwards with respect to discs; any given galaxy then has a $j_\ast$ that can be expressed as a linear superposition of a disc and a spheroid (a bulge). 

The resemblance of our $j_\ast-M_\ast-f_{\rm gas}$ plane (where at fixed $M_\ast$ a different $f_{\rm gas}$ produces a shift in $j_\ast$) with the $\mathcal{B}_\ast$ relation is clear. Both relations are valid for a variety of morphological types and are preserved along a broad mass span, and with a dependence of $j_\ast$ on $M_\ast$ with a slope of 2/3. The similarities are not unexpected since gas-poor galaxies usually have high $\mathcal{B}_\ast$, though the $f_{\rm gas}-\mathcal{B_\ast}$ relation is highly scattered. The above suggests that these two relations may be manifestations of a common physical mechanism. Finally, we note that the scatter is better quantified in the $j_\ast-M_\ast-f_{\rm gas}$ relation with respect to the $\mathcal{B}_\ast$ relation given that the uncertainties in $\mathcal{B}_\ast$ are difficult to estimate (\citealt{salo2015}; \citetalias{fall2018}).

\begin{figure}
    \centering
    \includegraphics[scale=0.45]{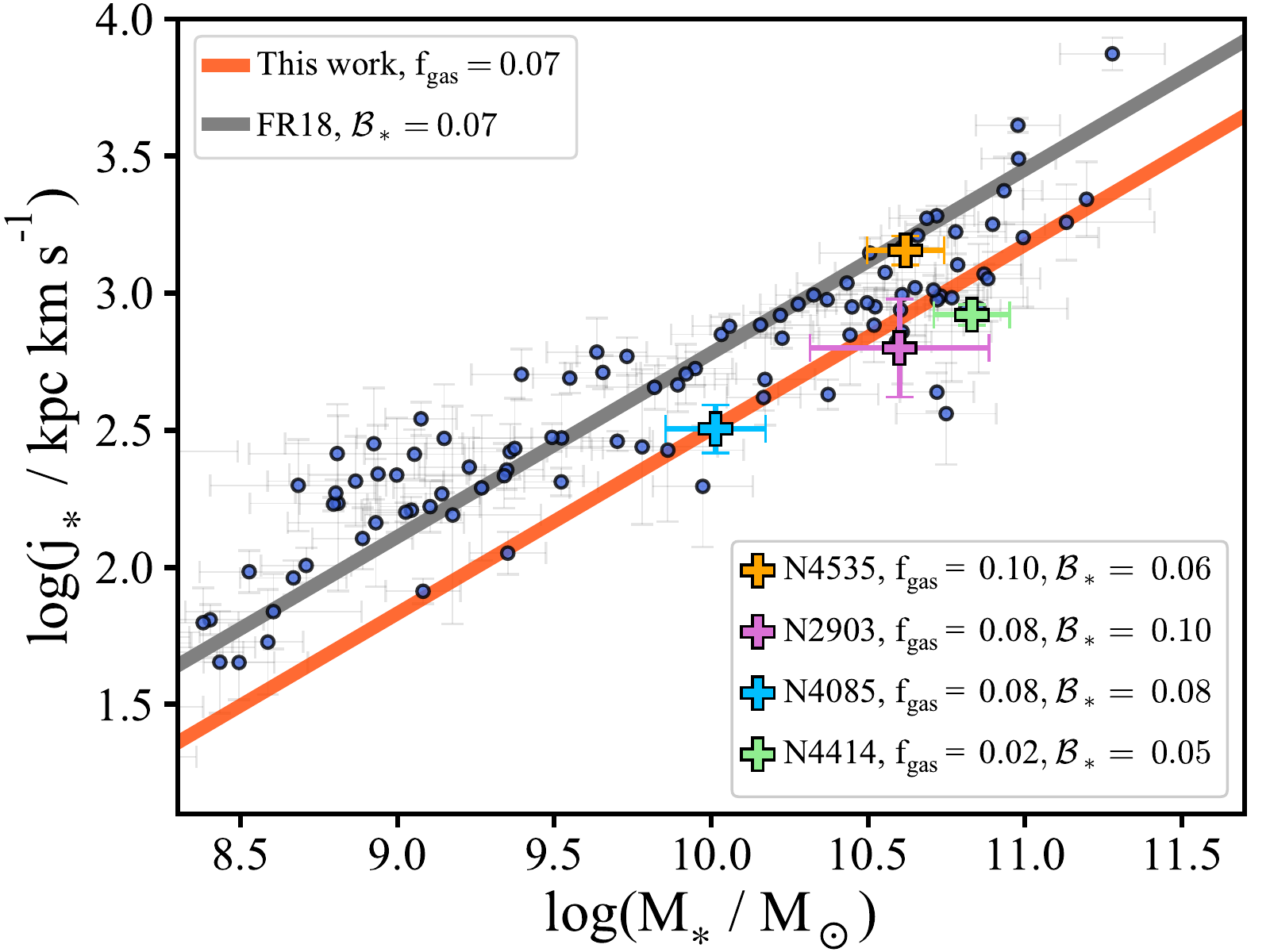}
    \caption{$j_\ast-M_\ast$ relation. Crosses show our galaxies with ($f_{\rm gas}, \mathcal{B}_\ast)~\leq~0.1$. The black and red lines show the expectations from \citetalias{fall2018} and this work, respectively. Only a fraction of our range in $M_\ast$ is shown.}
    \label{fig:fgasvsBT}
\end{figure}

Interestingly, there is a regime in which the $f_{\rm gas}$ relation makes a different prediction from the $\mathcal{B}_\ast$ relation. For galaxies that are almost gas-free and almost bulge-less (we note that galaxies with $\mathcal{B}_\ast \lesssim 0.1-0.2$ host pseudo-bulges rather than classical bulges; see Fig.~3 in \citetalias{fall2018}), the $f_{\rm gas}$ relation expects them to have a lower-than-average $j_{\rm \ast}$, while the $\mathcal{B}_\ast$ relation predicts a higher-than-average $j_\ast$. We tested this in Fig.~\ref{fig:fgasvsBT} by looking at the four galaxies in our sample with $f_{\rm gas} \leq 0.1$ and $\mathcal{B}_\ast \leq 0.1$ \citep{salo2015}. We also plot the expected lines given the average $\mathcal{B}_\ast$ and $f_{\rm gas}$ for these four galaxies. The points lie close to the $f_{\rm gas}$ relation and deviate from the $\mathcal{B}_\ast$ one. 
However, the evidence is not compelling given the low-number statistics. Finally, it is also important to mention that our galaxy sample is fairly different from that of \citetalias{fall2018}, with many more gas-dominated discs but a lack of early-type galaxies. These potential differences can be further explored with larger and more complete samples.

\subsection{The origin of the baryonic relation}
We then focused our attention on the $j_{\rm bar}-M_{\rm bar}-f_{\rm gas}$ relation. Its origin is likely related to different galaxy formation processes, such as variations in the angular momentum of the dark matter halos, selective gas accretion within the discs, and different gas accretion histories from the intergalactic medium \citep{fall83,postijstar,posti_spin,stevens2018,zoldan18}. Evolutionary processes such as stellar and active galactic nucleus feedback, mergers, and angular momentum transfer between galaxies and their dark matter halos are arguably also important (\citealt{leroy,duttonvandenbosch,romanowsky,lagos_eagle,zoldan18}). Still, while a complex interplay between all these processes is expected, it all results in the tight $j_{\rm bar}-M_{\rm bar}-f_{\rm gas}$ law we observe. Therefore, it is interesting to check whether or not simple mechanisms are able to capture the dominant processes that give rise to the $j_{\rm bar}-M_{\rm bar}-f_{\rm gas}$ relation.

\subsubsection{Disc instability}
We considered two models that have been proposed to explain the $j_{\rm bar}-f_{\rm gas}$ connection as a consequence of gravitational instability. First, \citet{obreschkow2016} proposed a model that relates $f_{\rm gas}$ with $j_{\rm bar}$ via
\begin{equation}
    f_{\rm{gas}} = \rm{min} \{1, 2.5 \textit{q}^{1.12}\},~~ \textit{q} = \textit{j}_{\rm bar} \sigma_{\rm gas} / \textit{G M}_{\rm bar}
    \label{eq:qparam}
,\end{equation}
with $q$ a stability parameter, $\sigma_{\rm gas}$ the gas velocity dispersion, and $G$ the gravitational constant. Deviations from Eq.~\ref{eq:qparam} may occur depending on the galaxy rotation curve, but they are expected to be small. These results were derived under a number of simplifying assumptions, but less idealised semi-analytic models are found to be in good agreement \citep{stevens2018}. From Eq.~\ref{eq:qparam}, one has $\log(f_{\rm gas})\propto~1.12[\log(j_{\rm bar})-~\log(M_{\rm bar})]$ and $j_{\rm bar} \propto M_{\rm bar}$ (this at fixed $f_{\rm gas}$ and assuming a constant $\sigma_{\rm gas}$). These dependences disagree with our best fitting plane, which has $\log(f_{\rm gas}) \propto 2.17\log(j_{\rm bar}) - 1.59\log(M_{\rm bar})$ and $j_{\rm bar} \propto M_{\rm bar}^{0.73}$ at fixed $f_{\rm gas}$.
Projecting our baryonic plane into the $f_{\rm gas}-q$ diagram shows that galaxies of a given $M_{\rm bar}$ follow parallel sequences of the form $f_{\rm gas} \propto q^{1/\beta_{\rm bar}} = q^{2.22}$, instead of $f_{\rm gas} \propto  q^{1.12}$. \footnote{We note that assuming a non-constant $\sigma_{\rm gas}$ is not enough to alleviate the mentioned discrepancies. To match our relations, $\sigma_{\rm gas} \propto M_{\rm bar}^{\sim 0.25}$ is required; however, it is observed that $\sigma_{\rm gas} \propto M_{\rm bar}^{0.07}$ (e.g. \citealt{murugeshan}).}

Also based on disc instability, \citet{romeo2020} proposed a set of scaling relations of the form $j_i \hat{\sigma}_i / (G M_i) \approx 1$, with $i$ denoting stars or gas and $\hat{\sigma}$ a mass-weighted radial average of the velocity dispersion $\sigma$. This relation produces the scaling $j_\ast \propto M_\ast^{0.5}$ (for $\hat{\sigma}_\ast \propto M_\ast^{0.5}$, as proposed by \citealt{romeo2020}) and $j_{\rm gas} \propto M_{\rm gas}$, very similar to the values found in \citetalias{paperIBFR} for the 2D relations. To incorporate $f_{\rm gas}$, we rewrote the above expression (using $M_i = M_{\rm gas} = f_{\rm gas} M_{\rm bar}$) as
\begin{equation}
\label{eq:romeo}
    \dfrac{j_{\rm gas}}{M_{\rm bar}} \dfrac{\hat{\sigma}_{\rm gas}}{G} = f_{\rm gas}~.
\end{equation}
Assuming a constant $\hat{\sigma}_{\rm gas}$, as in \citet{romeo2020}, this relation predicts $j_{\rm gas} \propto M_{\rm bar}$ at fixed $f_{\rm gas}$ and $j_{\rm gas} \propto f_{\rm gas}$ at fixed $M_{\rm bar}$. Instead, a corollary of our gas relation is that $j_{\rm gas} \propto M_{\rm bar}^{0.78}$ at fixed $f_{\rm gas}$ and that $j_{\rm gas} \propto f_{\rm gas}^{0.27}$ at fixed $M_{\rm bar}$. Therefore, the relation from \citet{romeo2020} also seems to depart from our data.

\subsubsection{Star formation efficiency}
\label{sec:SFE}
A more general possibility is that the link between $j_{\rm bar}$, $M_{\rm bar}$, and $f_{\rm gas}$ is related to the star formation efficiency in galaxies, as we briefly discuss here. We started by noting that at fixed $M_{\rm bar}$ the larger the $j_{\rm bar}$ of a galaxy is, the more extended its baryonic distribution will be.\footnote{Second-order effects related to the concentration of the host halo might also be relevant in determining the galaxy sizes \citep{posti_galaxyhalo}.} Also, it is well established that gas located in the outskirts of discs forms stars less efficiently than gas closer to the centre (e.g. \citealt{leroy}). Thus, at fixed $M_{\rm bar}$, a galaxy with a large $j_{\rm bar}$ also has a large $f_{\rm gas}$ since a large portion of its mass is located in the less star-forming outer regions. Qualitatively, this is in agreement with the fact that for our entire galaxy sample $j_{\rm gas}/j_\ast \approx R_{\rm gas}/R_\ast > 1$ (see Sect.~\ref{sec:similarslopes}). All this suggests that the connection between $j_{\rm bar}$ and $f_{\rm gas}$ may be a reflection of the mechanism responsible for a radially declining star formation efficiency and a radially increasing $f_{\rm gas}$ in galaxy discs (e.g. \citealt{leroy,krumholz2011,ceciVSFL}), but exploring this idea quantitatively (e.g. by investigating why the $j_{\rm gas}/j_\ast$ ratio is largely independent of $M_{\rm bar}$ and $f_{\rm gas}$; see Fig.~\ref{fig:ratio} but also Fig.~\ref{fig:RgasRstar}) is beyond the scope of the present letter.



\section{Conclusions}
\label{sec:conclusion}
In this letter we have used a high-quality sample of disc galaxies to study the relation between their specific angular momenta ($j$), masses ($M$), and gas fraction ($f_{\rm gas}$). The position of our galaxies in the $(j,M,f_{\rm gas}$) spaces can be described with planes such that galaxies with different $f_{\rm gas}$ follow parallel lines in the projected 2D $(j,M)$ spaces. Remarkably, our planes are preserved along a wide range of mass and morphology with very small ($\leq$ 0.1~dex) intrinsic scatter, which places the relations amongst the tightest scaling laws for disc galaxies. The $j_{\rm bar}-M_{\rm bar}-f_{\rm gas}$ plane is arguably the most fundamental, and it is even followed by populations of galaxies with extreme size, mass, and gas content, some of which were previously claimed to be outliers of the 2D $j-M$ relations.

The $j_\ast-M_\ast-f_{\rm gas}$ relation shows analogies with the $j_\ast-M_\ast-\mathcal{B}_\ast$ relation ($\mathcal{B}_\ast$ being the bulge-to-total mass fraction) previously discussed in the literature. Galaxies with fixed $f_{\rm gas}$ or $\mathcal{B_\ast}$ follow parallel relations of the form $j_\ast \propto M_\ast^{2/3}$. Most galaxies are well described by both the $f_{\rm gas}$ and $\mathcal{B}_\ast$ relations, while some discrepancies appear when looking at galaxies with low $f_{\rm gas}$ and low $\mathcal{B}_\ast$.

Finally, we show that models based purely on disc instability do not quantitatively reproduce the observed $j-M-f_{\rm gas}$ relations. We argue that the origin and behaviour of the $j_{\rm bar}-M_{\rm bar}-f_{\rm gas}$ law is closely related to the spatial distribution of gas and stars within galaxies, as well as to the radial variations in the star formation efficiency.

We stress that our relations offer a unique possibility to quantitatively test a variety of models, including hydrodynamical simulations and semi-analytic models, providing a powerful benchmark for theories on the formation of galactic discs. The slopes and intrinsic scatter of our $j-M-f_{\rm gas}$ planes are important requirements that hydrodynamical simulations and (semi-)analytic models should aim to reproduce.

\begin{acknowledgements}
We would like to thank our referee, Danail Obreschkow, for a very constructive and valuable report, and for his kind assistance regarding the use of \textsc{hyper-fit}. We thank Enrico di Teodoro for making available the data on `super spiral' galaxies for us. P.E.M.P., and F.F. are supported by the Netherlands Research School for Astronomy (Nederlandse Onderzoekschool voor Astronomie, NOVA), Project 10.1.5.6. P.E.M.P. benefited from individual comments and discussions with Andrea Afruni and Fernanda Rom\'an-Oliveira. L.P. acknowledges support from the Centre National d'\'{E}tudes Spatiales (CNES) and from the European Research Council (ERC) under the European Unions Horizon 2020 research and innovation program (grant agreement No. 834148). G.P. acknowledges support from NOVA, Project 10.1.5.18. E.A.K.A. is supported by the WISE research programme, which is financed by the Netherlands Organization for Scientific Research (NWO). We used SIMBAD and ADS services, as well the Python packages NumPy \citep{numpy}, Matplotlib \citep{matplotlib}, and SciPy \citep{scipy}, for which we are thankful.
\end{acknowledgements}

\begin{appendix}
\section{The role of molecular gas}
\label{sec:appH2}
In this letter we have assumed $M_{\rm gas} = 1.33M_{\rm HI}$ and $j_{\rm gas} = j_{\rm HI}$, neglecting any contribution from molecular (H$_2$) and ionised gas. Here we show that observationally motivated corrections to account for the presence of H$_2$ do not change our results.

To account for $M_{\rm H_2}$ in our estimates of $M_{\rm gas}$, we relied on the results from \citet{catinella2018}, who provide measurements of the ratio $M_{\rm H_2}/M_{\rm HI}$ as a function of $M_\ast$ for a large sample of nearby galaxies. We fitted a linear relation to their binned measurements (see their Fig.~9 and Table~3), finding
\begin{equation}
    \log(M_{\rm H_2}/M_\odot) = 0.26\log(M_\ast/M_\odot) + \log(M_{\rm HI}/M_\odot) - 3.4.
\end{equation}
The scatter is large, and we adopted an uncertainty of 50\% in $M_{\rm H_2}$. With this, we redefined $M_{\rm gas} = 1.33(M_{\rm HI} + M_{\rm H_2})$ and updated $M_{\rm bar}$ and $f_{\rm gas}$ accordingly.
For massive discs, the correction to $M_{\rm gas}$ is about 0.1~dex, which is smaller than the typical uncertainty in $M_{\rm gas}$ ($\sim$0.13~dex); the correction is even smaller for the dwarfs. The change in $f_{\rm gas}$ is of the same order, also being negligible at low masses and changing by up to 0.1~dex at the high-mass end; this change is also of the order of the typical uncertainty in $f_{\rm gas}$. Correspondingly, $M_{\rm bar}$ remains largely unchanged since the correction is smaller than 0.04~dex at all masses.\\

\noindent
Including H$_2$ can also affect $j_{\rm gas}$, as seen from the equation
\begin{equation}
    j_{\rm gas} = f_{\rm atm}j_{\rm HI} + (1-f_{\rm atm})j_{\rm H_2}~,
\end{equation}
where $j_{\rm HI}$ and $j_{\rm H_2}$ are the specific angular momenta of the atomic and molecular gas components, respectively, and $f_{\rm atm}$ is the atomic-to-total gas ratio, $f_{\rm atm} = M_{\rm HI}/(M_{\rm HI} + M_{\rm H_2})$.

\citet{OG14} provide measurements of $j_{\rm HI}$, $j_{\rm H_2}$, and $j_{\rm gas}$ for a sample of 16 spiral galaxies. In addition to this, we computed $j_{\rm HI}$ for four galaxies in our sample that have surface densities and CO rotation curves available from \citet{ceci_turbulence}. The typical ratio between $j_{\rm gas}$ and $j_{\rm HI}$ is 0.85, which translates into a shift of 0.07~dex. Thus, on average, including H$_2$ implies a correction to $j_{\rm HI}$ such that $\log(j_{\rm gas}) = \log(j_{\rm HI}) - 0.07$. The correction is of the same order as the average uncertainty in $j_{\rm gas}$, 0.08~dex.

We again fitted the 3D relations taking into account the above corrections to $M_{\rm gas}$, $f_{\rm gas}$, $j_{\rm gas}$, and $M_{\rm bar}$. The new coefficients for stars, gas, and baryons are listed in Table~\ref{tab:param_with_H2}. As expected, they are fully consistent with those reported in Table~\ref{tab:coeff_planes} within the uncertainties. Therefore, we conclude that including H$_2$ does not have a significant effect on the derivation of the $j-M-f_{\rm gas}$ laws, and our results remain unchanged.

\begin{table}[h]
\caption{Same as Table~\ref{tab:coeff_planes} but taking into account the presence of molecular gas.}
\label{tab:param_with_H2}
\begin{center} 
\begin{tabular}{ccccc}
        \hline \noalign{\smallskip}
 & $\alpha$ & $\beta$ & $\gamma$ & $\sigma_\perp$ \\ \noalign{\smallskip}
   \hline \noalign{\smallskip}
   Stars & 0.67 $\pm$ 0.03 & 0.55 $\pm$ 0.08 & -3.57 $\pm$ 0.23 & 0.11 $\pm$ 0.01 \\ \noalign{\smallskip}
    Gas & 0.75 $\pm$ 0.03 & -0.50 $\pm$ 0.04 & -4.46 $\pm$ 0.25 & 0.08 $\pm$ 0.01 \\ \noalign{\smallskip}
    Baryons & 0.72 $\pm$ 0.02 & 0.42 $\pm$ 0.05 & -4.23 $\pm$ 0.19 & $0.08 \pm 0.01$ \\ \noalign{\smallskip}
    \hline
\end{tabular}
  \end{center}
\end{table}

\newpage
\section{The $j_{\rm gas}/j_\ast$ ratio from our best fitting planes}
\label{sec:app_ratio}

\begin{figure}[h]
    \centering
    \includegraphics[scale=0.5]{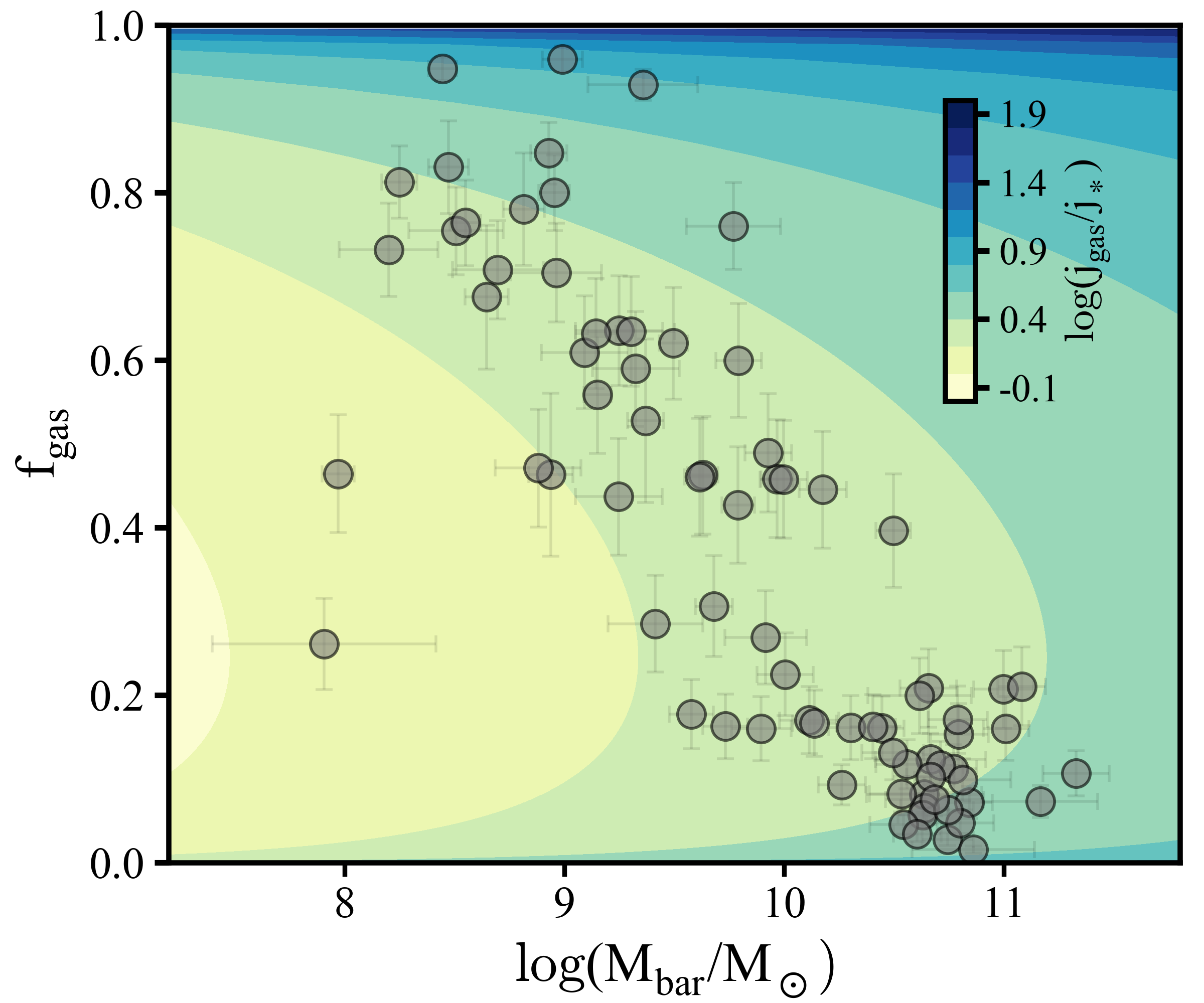}
    \caption{Relation between $M_{\rm bar}$, $f_{\rm gas}$, and the best fitting $j_{\rm gas}/j_\ast$ ratio according to Eq.~\ref{eq:RgasRstar}. The background shows increasing levels of $j_{\rm gas}/j_\ast$, and the grey points show our galaxies with convergent measurements of $j_{\rm gas}$ and $j_\ast$.}
      \label{fig:RgasRstar}
\end{figure}

The $j_{\rm gas}/j_\ast$ ratio can be obtained directly from our individual galaxy measurements, as shown in Fig.~\ref{fig:ratio}. Nevertheless, $j_{\rm gas}/j_\ast$ can also be obtained by using our best fitting stellar and gas relations. This allows us to extrapolate $j_{\rm gas}/j_\ast$ to values of $f_{\rm gas}$ and $M_{\rm bar}$ beyond our observations and, in principle, to neglect the observational uncertainties since they are accounted for in our best fitting planes. Using Eq.~\ref{eq:3dplanes} the ratio becomes
\begin{multline}
\label{eq:RgasRstar}
\log(j_{\rm gas}/j_\ast) = \alpha_{\rm gas}\log(f_{\rm gas} M_{\rm bar}) - \alpha_{\ast}\log[(1-f_{\rm gas}) M_{\rm bar}] \\ + (\beta_{\rm gas}-\beta_{\ast})\log(f_{\rm gas}) + \gamma_{\rm gas}-\gamma_{\ast}~,
\end{multline}
\noindent
and the corresponding surface according to Table~\ref{tab:coeff_planes} is shown in Fig.~\ref{fig:RgasRstar}. As can be seen, most of our galaxies lie within a region where $j_{\rm gas}/j_\ast \sim 2$. It will be interesting to see where other large samples of galaxies would lie in Fig.~\ref{fig:RgasRstar} and whether they also follow the expected dependence of $j_{\rm gas}/j_\ast$ on $M_{\rm bar}$ and $f_{\rm gas}$.

\end{appendix}

%
   \bibliographystyle{aa} 
   \bibliography{references.bib} 
%

\end{document}